\documentclass[aps,prl,superscriptaddress,twocolumn]{revtex4-1}

\usepackage{graphicx,graphics,epsfig}   
\usepackage{dcolumn}    
\usepackage{bm}         
\usepackage{amsmath}    
\usepackage{verbatim}   
\usepackage{color}      
\usepackage{subfigure}  
\usepackage{amsmath,amsfonts,amssymb,graphics,graphics,color,bbm}
\usepackage[rgb]{xcolor}
\usepackage{enumerate}
 \usepackage{amssymb}
 \usepackage{mathdots}
 \usepackage{amsthm}

\makeatletter
\def\Ddots{\mathinner{\mkern2mu
\raise8\p@\hbox{ . }\mkern2mu
\raise4\p@\hbox{ . }\mkern2mu\raise0\p@\hbox{ . }\mkern1mu}}
\makeatother

\newcommand{\sket}[1]{{\ensuremath{\lvert#1\rangle}}}
\newcommand{\lket}[1]{{\ensuremath{\left\lvert#1\right\rangle}}}
\newcommand{\ket}[1]{\if@display\lket{#1}\else\sket{#1}\fi}

\newcommand{\sbra}[1]{{\ensuremath{\langle#1\rvert}}}
\newcommand{\lbra}[1]{{\ensuremath{\left\langle#1\right\rvert}}}
\newcommand{\bra}[1]{\if@display\lbra{#1}\else\sbra{#1}\fi}

\newcommand{\sbraket}[2]{{\ensuremath{\langle#1\rvert#2\rangle}}}
\newcommand{\lbraket}[2]{{\ensuremath{\left\langle#1\!\left\rvert\vphantom{#1}#2\right.\!\right\rangle}}}
\newcommand{\braket}[2]{\if@display\lbraket{#1}{#2}\else\sbraket{#1}{#2}\fi}

\newcommand{\sketbra}[2]{{\ensuremath{\lvert #1\rangle\!\langle #2\rvert}}}
\newcommand{\lketbra}[2]{{\ensuremath{\left\lvert #1\right\rangle\!\!\left\langle #2\right\rvert}}}
\newcommand{\ketbra}[2]{\if@display\lketbra{#1}{#2}\else\sketbra{#1}{#2}\fi}

\newcommand{\proj}[1]{\ketbra{#1}{#1}}

\DeclareMathOperator{\Tr}{Tr}

\newtheorem{lemma}{Lemma}

\newcommand{\ba}{\begin{eqnarray}}
\newcommand{\ea}{\end{eqnarray}}
\newcommand{\ban}{\begin{eqnarray*}}
\newcommand{\ean}{\end{eqnarray*}}

\begin{document}

\title{Genuinely multipartite entangled quantum states with fully local hidden variable models and hidden multipartite nonlocality}

\author{Joseph Bowles}
\affiliation{D\'epartement de Physique Th\'eorique, Universit\'e de Gen\`eve, 1211 Gen\`eve, Switzerland}

\author{J\'er\'emie Francfort}
\affiliation{D\'epartement de Physique Th\'eorique, Universit\'e de Gen\`eve, 1211 Gen\`eve, Switzerland}

\author{Mathieu Fillettaz}
\affiliation{D\'epartement de Physique Th\'eorique, Universit\'e de Gen\`eve, 1211 Gen\`eve, Switzerland}

\author{Flavien Hirsch}
\affiliation{D\'epartement de Physique Th\'eorique, Universit\'e de Gen\`eve, 1211 Gen\`eve, Switzerland}

\author{Nicolas Brunner}
\affiliation{D\'epartement de Physique Th\'eorique, Universit\'e de Gen\`eve, 1211 Gen\`eve, Switzerland}

\date{\today}  

\begin{abstract}
The relation between entanglement and nonlocality is discussed in the case of multipartite quantum systems. We show that, for any number of parties, there exist genuinely multipartite entangled states which admit a fully local hidden variable model, i.e. where all parties are separated. Hence, although these states are entangled in the strongest possible sense, they cannot lead to Bell inequality violation considering general non-sequential local measurements. Then, we show that the nonlocality of these states can nevertheless be activated using sequences of local measurements, thus revealing genuine multipartite hidden nonlocality. 
\end{abstract}

\maketitle

The relation between quantum entanglement and nonlocality has been studied extensively in recent years; see e.g. \cite{review,valerio}. While both notions turn out to be equivalent for pure states \cite{gisin91,PR92}, the case of mixed state is still not understood. This is nevertheless desirable given the importance of entanglement and nonlocality from the point of view of the foundations of quantum theory and for quantum information processing \cite{review}.

This research was initiated by Werner \cite{werner89}, who presented a class of bipartite entangled states admitting a local hidden variable (LHV) model. This proved that the correlations obtained by performing arbitrary local projective measurements on such states can be perfectly simulated by a LHV model, hence using only classical resources. This was later extended to general non-sequential measurements, i.e. POVMs \cite{barrett02}. Since such states cannot lead to Bell inequality violation \cite{bell64}, they are referred to as `local' entangled states \cite{augusiak_review}. 

It turns out however that certain local entangled states can nevertheless lead to nonlocality when a sequence of local measurements is performed \cite{popescu95}. That is, the use of local filters can help to reveal (or activate) the nonlocality of the entangled state. This phenomenon, termed `hidden nonlocality', occurs even for entangled states admitting a LHV model for POVMs \cite{hirsch}. Other works showed that the nonlocality of local entangled states can be activated by performing joint measurements on several copies of the state \cite{palazuelos,cavalcanti}, or by placing many copies of the state in a quantum network \cite{dani,sen}.

Whereas the above questions have been intensively discussed for bipartite states, the relation between entanglement and nonlocality for multipartite systems is almost unexplored so far. Here one should nevertheless expect interesting and novel phenomenona, due to the rich structure of multipartite entanglement. In particular, there is a hierarchy of different forms of entanglement in multipartite systems, the strongest of which is genuine multipartite entanglement (GME). Similarly, the notion of genuine multipartite nonlocality (GMNL) has been discussed \cite{svetlichny,bancal13,gallego}, which represents the strongest form of nonlocality for multipartite systems. A first natural question is then whether there exist GME states, the correlations of which can be simulated by a LHV model. This was first discussed by T\'oth and Ac\'in \cite{toth05}, who presented a GME state of 3 qubits admitting a LHV model, but could not extend their construction to more parties. More recently, Augusiak et al.\cite{augusiak14} showed the existence of GME states of any number of parites which cannot lead to GMNL. Specifically, the authors discussed a class of GME states of $N$ parties, and constructed a LHV model in which the parties are separated into two groups. However, this model is essentially bipartite, as the $N$ parties cannot be completely separated. Beyond these few exploratory works, nothing is known to the best of our knowledge.

Here we report progress in understanding the relation between GME and nonlocality. First, we present a general technique for constructing multipartite entangled states admitting a fully LHV model, i.e. where all parties are separated. This allows us to show that there exist GME states of an arbitrary number of systems, which admit a fully LHV model for arbitrary POVM measurements. Moreover, we show that the nonlocality of these states can be activated using sequential measurements. Notably, the use of local filters allows us to obtain GMNL. To summarize, there exists multipartite states, entangled in the strongest possible sense, which do not exhibit even the weakest form of nonlocality when considering non-sequential measurements. However, when using sequences of measurements, the strongest form of multipartite nonlocality can be obtained. We conclude with a series of open questions.

\section{Preliminaries} 
\emph{Genuine multipartite entanglement.}---Consider $N$ parties sharing a multipartite quantum state $\rho$ acting on ${\cal{H}}_1\otimes\cdots\otimes{\cal{H}}_N$, where ${\cal{H}}_i$ is the local Hilbert space of party $i$. Denote by $(b,\bar{b})\in {\cal{B}}$ a bipartition of the $N$ parties. If $\rho$ can be decomposed as a mixture of states that are each separable on some bipartition of the Hilbert space then we have
\begin{align}\label{GME}
\rho=\sum_{(b,\bar{b})\in{\cal{B}}} p_b\left( \sum_{j} q^{b}_j \proj{\Phi_j}_b\otimes\proj{\Phi_j}_{\bar{b}}\right),
\end{align}
with $\sum_{b}p_b=\sum_{j} q_j^b=1$ and $\proj{\Phi_j}_b$ acts on the Hilbert space specified by the partition $b$ (and similarly for $\proj{\Phi_j}_{\bar{b}}$). If $\rho$ does not admit such a decomposition then it is GME. Such states can thus not be created via LOCC using only biseparable states. 

Determining whether a given state is GME is challenging, as one must search over all possible decompositions \eqref{GME}. However, there are sufficient conditions for an $N$-qubit state to be GME \cite{guhne,huber1,hao}. Write the state $\rho$ in the canonical basis $\ket{0,0,\cdots,0},\ket{0,0,\cdots,1},\cdots,\ket{1,1,\cdots,1}$ as 
\small\begin{align}
\rho=\left(  \begin{array}{cccccccc}
    c_{1} & ¥ & ¥ & ¥ & ¥ & ¥ & ¥ & z_{1} \\ 
    ¥ & c_{2} & ¥  & ¥ & ¥ & ¥ & z_{2} & ¥ \\ 
    ¥ & ¥ & \ddots & ¥ & ¥ & \iddots & ¥ & ¥ \\ 
    ¥ & ¥ & ¥ & c_{n} & z_{n} & ¥ & ¥ & ¥ \\ 
    ¥ & ¥ & ¥ & z_{n}^{*} & d_{n} & ¥ & ¥ & ¥ \\ 
    ¥ & ¥ & \iddots & ¥ & ¥ & \ddots & ¥ & ¥ \\ 
    ¥ & z_{2}^{*} & ¥ & ¥ & ¥ & ¥ & d_{2} & ¥ \\ 
    z_{1}^{*} & ¥ & ¥ & ¥ & ¥ & ¥ & ¥ & d_{1} \\ 
  \end{array}
\right)
\end{align}\normalsize
(we only write the elements of interest), where $n=2^{N-1}$. Then $\rho$ is GME if 
\begin{align}\label{gmecon}
C(\rho)=2\max_i\{|z_{i}|-w_{i}\}>0,
\end{align}
where $w_{i}=\sum_{j\neq i}^{n}\sqrt{c_{j} d_{j}}$. The value of $C(\rho)$ gives a lower bound for the genuine multipartite concurrence of $\rho$, and becomes exact for the case of so called qubit X-matrices \cite{heshemi}.

\emph{Non-locality}---Consider again the state $\rho$, where now each party can make measurements labelled $x_{i}$ obtaining outcomes $a_i$, specified by the measurement operators $M_{a_i|x_i}$, with $M_{a_i|x_i}\geq0$ and $\sum_{a_i}M_{a_i|x_i}=\openone$. The probability to see the outputs $\mathbf{a}=(a_1,\cdots,a_N)$ given the inputs $\mathbf{x}=(x_1,\cdots,x_{N})$ is given by
\begin{align}
p(\mathbf{a}|\,\mathbf{x})&=\Tr\left[\rho\,(\otimes_{i=1}^{N}M_{a_i|x_i} )\right].
\end{align}
The state $\rho$ is called (fully) local if, for all possible measurement operators $M_{a_i|x_i}$, the statistics $p(\mathbf{a}|\,\mathbf{x})$ can be reproduced by a local hidden variable (LHV) model:
\begin{align}\label{local}
p(\mathbf{a}|\mathbf{x})=\int_{\lambda} q_{\lambda} p_{\lambda}(a_1|x_1)p_{\lambda}(a_2|x_2)\cdots p_{\lambda}(a_N|x_N) \text{d}\lambda,
\end{align}\normalsize
where $q_{\lambda}$ is a probability density over the shared variable $\lambda$ and the $p_{\lambda}(a_i|x_i)$'s are probability distributions, called local response functions. Likewise, if \eqref{local} cannot be satisfied then the state is said to be nonlocal, as witnessed by the violation of (some) Bell inequality. 

One may also consider a weaker notion of locality, whereby the correlations are not demanded to be local with respect to all parties (as in \eqref{local}), but instead to be (mixtures of) correlations that are each local across some bipartition. Again denoting by $(b,\bar{b})\in{\cal{B}}$ a bipartition of the parties, these correlations take the form 
\begin{align}\label{gmnl}
p(\mathbf{a}|\mathbf{x})=\sum_{(b,\bar{b})\in{\cal{B}}}p_{b}\int_{\lambda} q_{\lambda}^b p_{\lambda}(\mathbf{a}_{b}|\mathbf{x}_b)p_{\lambda}(\mathbf{a}_{\bar{b}}|\mathbf{x}_{\bar{b}})\text{d}\lambda,
\end{align}
where $\mathbf{a}_b$, $\mathbf{x}_b$ denote the inputs and outputs for the bipartition $b$. Note that \eqref{local} implies \eqref{gmnl} but not necessarily the converse. Correlations which cannot be written in the above form are called \emph{genuinely multipartite nonlocal} (GMNL) and represent the strongest form of multipartite nonlocality \cite{svetlichny}. Here, for simplicity, we put no restrictions on the probability distributions $p_{\lambda}(\mathbf{a}_{b}|\mathbf{x}_{b})$, $p_{\lambda}({\mathbf{a}}_{\bar{b}}|{\mathbf{x}}_{\bar{b}})$ other than positivity and normalisation (for example they may be signalling); note that more sophisticated definitions of GMNL were proposed \cite{gallego,bancal13}. The $N$-party GHZ state $\ket{\text{GHZ}}=(\ket{0}^{\otimes N}+\ket{1}^{\otimes N})/\sqrt{2}$ is known to produce correlations which are GMNL, as proven by the violation of the Svetlichny inequalities \cite{svetlichny,collins,seevinck}.

\emph{GME and nonlocality.}---The link between GME and nonlocality is almost unexplored so far. For $N=3$, Toth and Acin constructed a genuine tripartite entangled state admitting a fully LHV model (i.e. of the form \eqref{local}) for arbitrary local projective measurements \cite{toth05}. Recently, Augusiak et al. \cite{augusiak14} presented GME states of $N$ qubits which cannot lead to GMNL. More precisely, they constructed a LHV model for some bipartition of $N$ qubits, i.e. of the form \eqref{gmnl}. However, it is still unknown if there exist GME states which admit LHV models that are fully local, i.e. that satisfty \eqref{local}, for any possible measurements. This is what we show in the next section.


\begin{figure*}
\includegraphics[scale=0.89]{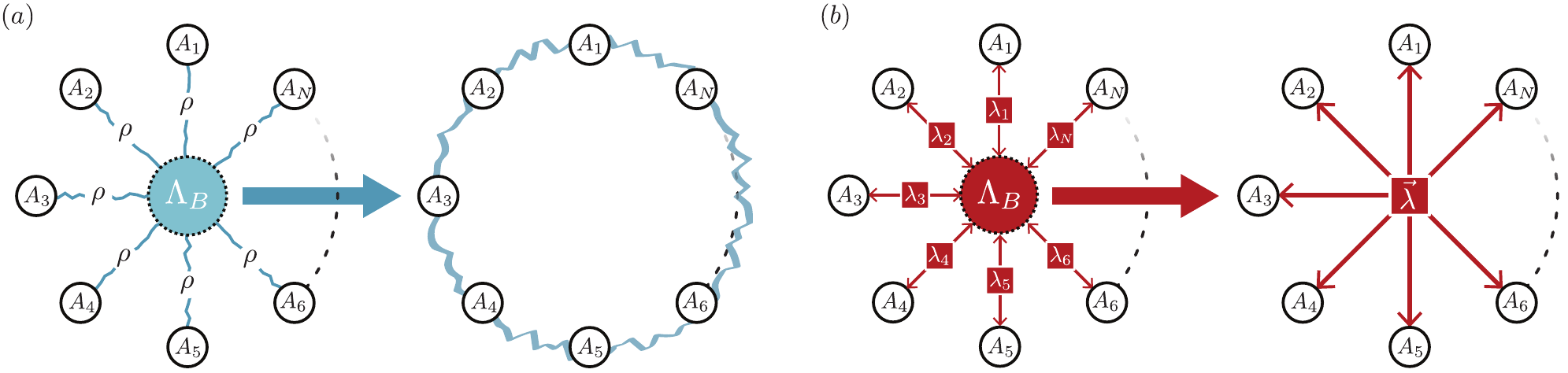}
\caption{\label{fig} Construction of multipartite states admitting a fully local model. (a) Construction of the state. First, place $N$ copies of a bipartite state $\rho$ in a star-shaped network. Then, apply a map $\Lambda_B$ at the central node (i.e. on parties $B_1 \cdots B_{N}$), and trace out these parties. We thus obtain an $N$-partite state, $\rho_{A_1\cdots A_{N}}$ (represented by the blue wiggly line), shared by parties $A_1 \cdots A_N$. (b) LHV model. If $\rho$ admits a LHS model, one can simulate the correlations of the star-shaped network for $\rho^{\otimes N}$, whereby the central node receives the hidden states $\sigma_{\lambda_i}$ independently from each source and the parties $A_i$ receive hidden variables $\lambda_i$. One may now correlate the individual $\lambda_{i}$'s by having the map $\Lambda_B$ act on the hidden states, i.e. we can define a new distribution over $\vec{\lambda}=(\lambda_1,\cdots,\lambda_N)$ that depends on $\Tr[\Lambda_{B}(\otimes_i\,\,\sigma_{\lambda_i})]$. If each party $A_i$ uses the same response function as in the LHS model for $\rho$, then the resulting statistics on parties $A_1\cdots A_N$ simulate exactly the state $\rho_{A_1\cdots A_{N}}$.}
\end{figure*}

 \section{Method}
Our main tool is a simple method to construct entangled $N$-party states which admit a LHV model. Specifically, we start by considering a bipartite entangled state $\rho$ which is `unsteerable', that is, which cannot be used to demonstrate steering. Formally, this means that $\rho$ admits a so-called local hidden state (LHS) model \cite{wiseman07}, hence its correlations can be decomposed as
%
\begin{align}\label{us}
p(ab|xy)&=\Tr[\rho\,M_{a|x}\otimes M_{b|y}] \nonumber\\
&=\int q_{\lambda}p_{\lambda}(a|x)\Tr[\sigma_{\lambda}M_{b|y}]\text{d}\lambda,
\end{align}
where $\sigma_{\lambda}$ is the local hidden state, distributed with density $q_{\lambda}$, and $B_{b|y}$ denotes Bob's measurement operator. Clearly, an unsteerable state is local (with $p(b|y,\lambda)=\Tr[\sigma_{\lambda} M_{b|y}]$), while the opposite may not hold in general. 

Next, we combine several copies of $\rho$ in a star-shaped network (see Fig.1). This allows one to construct a multipartite entangled state admitting a fully local model. Specifically, we have the following:

%
\begin{lemma}
Let $\rho$ be a quantum state acting on ${\cal{H}}_{A_1}\otimes{\cal{H}}_{B_1}$. The state $\rho^{\otimes N}$ therefore acts on ${\cal{H}}_{A_1}\otimes\cdots\otimes{\cal{H}}_{A_N}\otimes{\cal{H}}_{B_1}\otimes\cdots\otimes{\cal{H}}_{B_N}={\cal{H}}_{A}\otimes{\cal{H}}_{B}$. Furthermore let $\Lambda_{B}$ be a completely positive linear map acting on ${\cal{H}}_{B}$. If $\rho$ is unsteerable from $A_1$ to $B_1$, i.e. admits a decomposition \eqref{us}, then the $N$-party state 
\begin{align}\label{map}
\rho_{A_1\cdots A_N}=\frac{\Tr_{B}\left[\openone_{A}\otimes\Lambda_{B}(\rho^{\otimes N})\right]}{\Tr[\openone_{A}\otimes\Lambda_{B}(\rho^{\otimes N})]}
\end{align}
admits a local hidden variable model, of the form \eqref{local}, on the $N$-partition $A_1/A_2/\cdots /A_{N-1}/A_N$.
\end{lemma}

The intuition behind the above lemma is given in Fig.1. 
A complete proof is given in Appendix A. 

Note that we have not specified the class of local measurements for which the LHV model is valid in the above lemma. If $\rho$ has a LHS model for projective measurements, then $\rho_{A_1\cdots A_N}$ will have a LHV model for projective measurements; similarly for POVMs. Note also that one can generalise slightly the result of Lemma 1 (see Appendix A). Specifically, one can use different unsteerable states in each arm of the start-shaped network rather than the same state $N$ times, and one can choose not to perform the trace over $B$ and keep the centre party.


\section{GME states with fully local model}
We now use Lemma 1 to construct $N$-qubit states which admit a fully local model. We then prove these states to be GME for all $N$. Specifically, consider the class of two-qubit states  
\begin{align}\label{mafalda}
\rho_{\alpha,\theta}=\alpha \proj{\psi_{\theta}}+(1-\alpha)\rho_{A}^{\theta}\otimes\frac{\openone}{2},
\end{align}
where $0\leq \alpha \leq 1$, $0\leq \theta \leq \pi/4$, $\ket{\psi_{\theta}}=\cos\theta\ket{00}+\sin\theta\ket{11}$, and  $\rho_{A}^{\theta}=\Tr_{B}\ket{\psi_{\theta}}\bra{\psi_{\theta}}$. These states are entangled for all $\theta\in\,]0,\pi/4]$ if $\alpha> 1/3$. Furthermore, they are unsteerable from Alice to Bob for arbitrary projective measurements if the relation 
\begin{align}\label{uscon}
\cos^2 (2\theta) \geq \frac{2\alpha-1}{(2-\alpha)\alpha^3}
\end{align}
holds \cite{bowles15}. Hence, for any $0\leq\alpha<1$ one may find a corresponding $\theta>0$ such that $\rho_{\alpha,\theta}$ is unsteerable. We now define the completely positive linear map
\begin{align}
\Lambda_{B}(\sigma)=F_B \sigma F_B^{\dagger} \, ,   \,\, 
F_B=\ket{0}\left[\bra{0,0, \cdots ,0}+\bra{1,1,\cdots ,1}\right], \nonumber
\end{align}
which projects the systems of $B_1\cdots B_N$ onto a $N$-qubit GHZ state. We may now define the $N$-party state $\rho_{A_1\cdots A_N}$ by using $\rho_{\alpha,\theta}$ and $\Lambda_B$ in \eqref{map}. In Appendix B we show that the concurrence of this state for a fixed $N$, $\alpha$, $\theta$ is given by
\small\begin{align}\label{concurrence}
&C(\rho_{A_1\cdots A_N})= \frac{2\sin^N (2\theta) \left(\alpha^N+\left[\frac{1+\alpha}{2}\right]^N+\left[\frac{1-\alpha}{2}\right]^N-1\right)}{\left[1+\alpha\cos 2\theta\right]^N +\left[1-\alpha\cos 2\theta\right]^N}.
\end{align}\normalsize
It follows that for any $N$, one can find parameters $\alpha,\theta$ such that (i) condition \eqref{uscon} is satisfied (ensuring that $\rho_{\alpha,\theta}$ has a LHS model), and (ii) $C(\rho_{A_1\cdots A_N})>0$, proving that $\rho_{A_1\cdots A_N}$ is GME. To give a specific example, take $\alpha=1-1/N^2$ and $\theta>0$ such that \eqref{uscon} is saturated. One sees that the denominator of \eqref{concurrence} and $\sin^{N}2\theta$ are both positive. We therefore need
\small\begin{align}
&\alpha^N+\left[\frac{1+\alpha}{2}\right]^N+\left[\frac{1-\alpha}{2}\right]^N>1
\end{align}\normalsize
to be positive for all $N\geq 2$. For the case $N=2$ one has $\alpha=3/4$ and we find $43/32>1$. For $N>2$, upon substituting $\alpha=1-1/N^2$ the left hand side becomes
\small\begin{align}
\left[1-\frac{1}{N^2}\right]^N+\left[1-\frac{1}{2N^2}\right]^N+\left[\frac{1}{2N^2}\right]^N \nonumber\\
>2\left[1-\frac{1}{N^2}\right]^N> 2\left[1-\frac{1}{N}\right]> 1
\end{align}\normalsize
where for the first inequality use the fact that $[1-1/N^2]^N<[1-1/2N^2]^N$ and $[1/2N^2]^N>0$, and the second inequality follows from Bernoulli's inequality.

\emph{Extension to general measurements.}---A natural question is now to find a GME state with a fully local model, considering general POVMs. While the states $\rho_{\alpha,\theta}$ are not known to admit a LHS model for POVMs, we can nevertheless proceed differently. Starting from $\rho_{A_1\cdots A_N}$, we can in fact construct another state, $\rho_{\text{\tiny{GME}}}$, which is both GME and local for POVM measurements. 

Specifically, define $\rho_{A_{1}\cdots A_{k}}=\Tr_{A_{k+1} \cdots A_{N}}[\rho_{A_1\cdots A_N}]$ and denote by $\circlearrowleft[\rho]$ the unnormalised and symmetrised version of $\rho$. Then the state 
\begin{align}\label{povmlocal}
\rho_{\text{\tiny{GME}}}=\frac{1}{2^{N}}\left[\rho_{A_1\cdots A_N}+\sum_{j=0}^{N-1}\circlearrowleft\left[\rho_{A_{1}\cdots A_{j}}\otimes\ketbra{2}{2}^{\otimes N-j}\right]\right]
\end{align}
admits a fully local model, for arbitrary local POVMs. Note that $\proj{2}$ denotes the projector onto a subspace orthogonal to the qubit subpace. The above follows from a straightforward extension of Protocol 2 of Ref. \cite{hirsch} to the case of $N$ parties.

To conclude, we have to show that the state is GME. Note that if each party makes a local projection on the qubit subspace $\proj{0}+\proj{1}$ then the resulting (renormalised) state is $\rho_{A_1\cdots A_N}$, which is GME. Since one cannot create GME using stochastic local operations, it follows that $\rho_{\text{\tiny{GME}}}$ is GME.

\section{Hidden genuine multipartite nonlocality}
We showed that GME states can admit a fully LHV model for arbitrary non-sequential measurements. A natural question now is whether these states have hidden nonlocality \cite{popescu95}, that is, whether nonlocality could be revealed via sequences of measurements. A sufficient condition for the existence of hidden nonlocality is the possibility of transforming the initial state using local stochastic operations, i.e. local filters, to another state that violates some Bell inequality (see e.g. \cite{rodrigo}). Below we will see that the states $\rho_{\text{\tiny{GME}}}$ have genuine multipartite hidden nonlocality. Furthermore, the activation of nonlocality is maximal, in the sense that the filtered state exhibits GMNL, despite the initial state being fully local. 

Consider $N$ parties sharing $\rho_{\text{\tiny{GME}}}$. Let each party perform a local filtering operation given by 
\begin{align}
G_\epsilon=\epsilon \proj{0} + \proj{1},
\end{align}
hence transforming $\rho_{\text{\tiny{GME}}}$ to the state
\begin{align}
\rho_{\epsilon}=\frac{G_\epsilon^{\otimes N}\rho_{\text{\tiny{GME}}}\,G_\epsilon^{\otimes N}}{\Tr[ G_\epsilon^{\otimes N}\rho_{\text{\tiny{GME}}}\,G_\epsilon^{\otimes N}]}.
\end{align}
In Appendix C we prove that for $\epsilon=\tan \theta$ (where $\theta$ is the parameter in \eqref{mafalda}), the filtered states is essentially a pure $N$-party GHZ state $[\ket{0}^{\otimes N}+\ket{1}^{\otimes N}]/\sqrt{2}$. Specifically, the fidelity between the two states is given by
\begin{align}
&{\cal{F}}(\rho_{\epsilon},\proj{\text{GHZ}})=\bra{\text{GHZ}}\rho_{\epsilon}\ket{\text{GHZ}}\nonumber\\[2pt]
&\quad\quad=\frac{1}{2}\left[\alpha^N +\left(\frac{1+\alpha}{2}\right)^N+\left(\frac{1-\alpha}{2}\right)^N\right].
\end{align}
which tends to 1 when $\alpha$ is sufficiently close to $1$. Since the GHZ state is known to exhibit GMNL for any $N$, in particular via violation of the Svetlichny inequalities \cite{collins,seevinck} (which are robust to noise), it follows that $\rho_{\epsilon}$ can also be made GMNL.

\section{conclusion}
We showed that GME states can admit a fully LHV model, for any number of parties. This demonstrates a maximal inequivalence between multipartite entanglement and multipartite nonlocality for non-sequential measurements. Interestingly, this gap can disappear when sequential measurements are considered, and the strongest form of nonlocality can be activated, thus highlighting the relevance of sequential measurements in multipartite nonlocality. 

It would be interesting to find examples of local GME states with high generalized concurrence. In order to do so, the method we presented for constructing multipartite local entangled states could be further explored, in particular starting from different bipartite unsteerable states. Finally, by keeping the central node in the network, one can construct multipartite LHS models where one of the parties has a quantum response function, and hence may prove useful in the study of multipartite steering \cite{multisteer}.\newline\newline\newline



\emph{Acknowledgements.} We thank Marco T\'{u}lio Quintino for discussions. We acknowledge financial support from the Swiss National Science Foundation (grant PP00P2\_138917 and Starting grant DIAQ).

\section{appendix A: General method proof}
Here we give a complete proof of Lemma 1. We define the state $\rho_\Lambda$ as
\begin{align}\label{rholam}
\rho_\Lambda=\frac{\openone_A\otimes\Lambda_B(\rho^{\otimes N})}{{\cal{N}}},
\end{align}
where 
\ba {\cal{N}}=\Tr[\openone_A\otimes\Lambda_B(\rho^{\otimes N})]=\Tr[\Lambda_{B}(\rho_{B}^{\otimes N})]
\ea 
with $\rho_{B}=\Tr_A[\rho]$. Note that $\rho_{A_1\cdots A_N}=\Tr_{B}[\rho_{\Lambda}]$.

We first show that the $N+1$ party distribution 
\begin{align}\label{localstats}
p(\mathbf{a}b|\mathbf{x}y)&=\Tr\left[(\otimes_{i=1}^{N}M_{a_i|x_i})\otimes M_{b|y}\;\rho_{\Lambda}\right]
\end{align}
admits a LHV model on the $N+1$ partition $A_1/A_2/\cdots/A_N/B$. Note that since the parties $B_1\cdots B_N = B$ now form a single party, the operator $M_{b|y}$ acts on the Hilbert space ${\cal{H}}_{B_1}\otimes\cdots \otimes{\cal{H}}_{B_N}$ and may be entangled across this space. Since a LHV model for a state clearly implies a LHV model for any subsystems of that state,  proving a LHV model for $\rho_{\Lambda}$ then implies a LHV model for $\Tr_{B}[\rho_{\Lambda}]=\rho_{A_1\cdots A_N}$, proving Lemma 1. To this end, we show the existence of a shared variable $\vec{\lambda}$ with corresponding normalised probability density $Q(\vec{\lambda})$ and response functions for the $N+1$ parties such that the corresponding LHV model reproduces the statistics \eqref{localstats}.

Replacing $\rho_{\Lambda}$ by \eqref{rholam} and denoting the dual map of $\Lambda_B$ by $\Lambda^*_B$ we have
\small\begin{align}
p(\mathbf{a}b|\mathbf{x}y)&=\frac{1}{{\cal{N}}}\Tr\left[(\otimes_iM_{a_i|x_i})\otimes M_{b|y}\;\openone_A\otimes\Lambda_B(\rho^{\otimes N})\right] \nonumber \\
					&=\frac{1}{{\cal{N}}}\Tr\left[(\otimes_iM_{a_i|x_i})\otimes \Lambda^{*}_{B}(M_{b|y}) \rho^{\otimes N}\right]\nonumber \\
					&=\frac{1}{{\cal{N}}}\Tr\left[\Tr_{A}\left[(\otimes_iM_{a_i|x_i}) \rho^{\otimes N}\right]\Lambda^{*}_{B}(M_{b|y}) \right] \nonumber \\
					&=\frac{1}{{\cal{N}}}\Tr\left[\left(\otimes_i\Tr_{A_i}\left[M_{a_i|x_i}\otimes \openone \,\rho \right]\right)\Lambda^{*}_{B}(M_{b|y})\right].
\end{align}\normalsize
Since we assume the state $\rho$ to be unsteerable, it follows that (for examples see \cite{wiseman07})
\begin{align}\label{usass}
\Tr_{A_i}\left[M_{a_i|x_i}\otimes \openone\,\rho\right]=\int q_{\lambda_i} p_{\lambda_i}(a_i|x_i)\sigma_{\lambda_{i}}\text{d}\lambda_i.
\end{align}
Combining this with the above we have
\small\begin{align}\label{modelfin}
&p(\mathbf{a}b|\mathbf{x}y)\nonumber\\
&\;=\frac{1}{{\cal{N}}}\Tr\left[\left(\otimes_i\int_{\lambda_i} q_{\lambda_i}p_{\lambda_i}(a_i|x_i)\sigma_{\lambda_i}\text{d}\lambda_{i}\right)\Lambda^{*}_{B}(M_{b|y})\right] \nonumber \\
&\;=\int_{\lambda_1}\cdots\int_{\lambda_N}\frac{q_{\lambda_1}\cdots\,q_{\lambda_N}}{{\cal{N}}}p_{\lambda_1}(a_1|x_1)\cdots p_{\lambda_N}(a_N|x_n)\nonumber \\
&\quad\quad\quad\quad\quad\quad\times \Tr\bigg[\left(\otimes_i \sigma_{\lambda_i}\right)\Lambda^{*}_{B}(M_{b|y})\bigg]\text{d}\lambda_{1}\cdots\text{d}\lambda_{N}\nonumber \\
&\;=\int_{\vec{\lambda}}Q(\vec{\lambda})\,p_{\lambda_1}(a_1|x_1)\cdots p_{\lambda_N}(a_N|x_n)\Tr\left[\sigma_{\vec{\lambda}}M_{b|y}\right]\text{d}\vec{\lambda},
\end{align}\normalsize
where $\vec{\lambda}=(\lambda_1,\cdots,\lambda_N)$ and we have 
\begin{align}
\sigma_{\vec{\lambda}}&= \frac{\Lambda_{B}(\otimes_{i}\sigma_{\lambda_i})}{\Tr[\Lambda_{B}(\otimes_{i}\sigma_{\lambda_i})]}\,\,,\\
Q(\vec{\lambda})&=\frac{\prod_{i}q_i}{{\cal{N}}}\Tr[\Lambda_{B}(\otimes_{i}\sigma_{\lambda_i})].
\end{align}
Equation \eqref{modelfin} is now in the precise form of a LHV model. The shared variable consists of the vector $\vec{\lambda}=(\lambda_1,\cdots,\lambda_N)$ which is distributed to the $N$ parties with probability density $Q(\vec{\lambda})$. Conditioned on $\vec{\lambda}$, the response functions for parties $A_1 \cdots A_N$ remain unchanged whereas party $B$ outputs according to $p(b|y,\vec{\lambda})=\Tr[ \sigma_{\vec{\lambda}} M_{b|y}]$, which is a valid probability distribution since $\sigma_{\vec{\lambda}}$ is a normalised quantum state. Furthermore since $\Lambda_B$ is positive we have $Q(\vec{\lambda})>0$ and 
\begin{align}
\int_{\vec{\lambda}} Q(\vec{\lambda})\text{d}\vec{\lambda} &= \int \frac{\prod_i q_i}{{\cal{N}}} \Tr\left[ \Lambda_{B}(\otimes_{i}\sigma_{\lambda_i})\right]\text{d}\vec{\lambda} \nonumber\\
&=\frac{1}{{\cal{N}}}\Tr\left[\Lambda_{B}\left(\otimes_{i}\int_{\lambda_i}q_{\lambda_i}\sigma_{\lambda_i}\text{d}\lambda_{i}\right)\right]\nonumber\\
&=\frac{1}{{\cal{N}}}\Tr\left[\Lambda_{B}\left(\rho_{B}^{\otimes N}\right)\right]\nonumber\\
&=1,
\end{align}
where the third line follows from $\eqref{usass}$ by setting say $A_{1|x_i}=\openone$ and consequently $p(1|x_1,\lambda)=1$. Hence $Q(\vec{\lambda})$ is indeed a probability density.

\begin{appendix}
\section{Appendix B: Calculation of $\mathbf{C(\boldsymbol{\rho}_{A_1\cdots A_{N}})}$}
Here we give a detailed derivation of \eqref{concurrence}. We first write the state \eqref{mafalda} as
\begin{align}\label{rhoi}
\rho_{\alpha,\theta}&= \left[\frac{1+\alpha}{2}\right]\left(c^2\proj{00}+s^2\proj{11}\right)\nonumber\\
&\quad +\left[\frac{1-\alpha}{2}\right]\left(c^2\proj{01}+s^2\proj{10}\right)\nonumber\\
&\quad\quad\quad\quad\quad+\alpha c s\left(\ketbra{00}{11}+\ketbra{11}{00}\right),
\end{align}
where $c,s$ denote $\cos \theta$ and $\sin \theta$ respectively. To begin, we consider the unormalised state
\begin{align}\label{target}
\rho_F=\Tr_{B}\left[[\openone_A\otimes F_B]\,\rho_{\alpha,\theta}^{\otimes N}\,[\openone_A\otimes F_B]\right],
\end{align}
where
\begin{align}
F_B=\ketbra{0}{00\cdots 0} +\ketbra{0}{11\cdots 1}
\end{align}
acts on ${\cal{H}}_B$. Notice that $\rho_{A_1\cdots A_N}=\rho_F/\Tr[\rho_F]$ and so $C(\rho_{A_1\cdots A_N})=C(\rho_F)/\Tr[\rho_F]$. After performing the partial trace of \eqref{target} we obtain
 \begin{align}\label{fourterms}
 \rho_{F}=\quad&\openone_A\otimes\bra{00\cdots 0}_B\;\rho_{\alpha,\theta}^{\otimes N}\;\openone_A \otimes \ket{00\cdots 0}_B\nonumber \\
 +&\openone_A\otimes\bra{11\cdots 1}_B\;\rho_{\alpha,\theta}^{\otimes N}\;\openone_A \otimes \ket{11\cdots 1}_B \nonumber \\
 +&\openone_A\otimes\bra{00\cdots 0}_B\;\rho_{\alpha,\theta}^{\otimes N}\;\openone_A \otimes \ket{11\cdots 1}_B \nonumber \\
 +&\openone_A\otimes\bra{11\cdots 1}_B\;\rho_{\alpha,\theta}^{\otimes N}\;\openone_A \otimes \ket{00\cdots 0}_A.
 \end{align} 
We consider each of these four terms separately. For the first term, the only non-zero contributions coming from $\rho_{\alpha,\theta}^{\otimes N}$ will correspond the $N$-fold tensor product of combinations of the projectors $\proj{00}$ and $\proj{10}$ with their corresponding weights. Hence, this will contribute diagonal terms to $\rho_F$. For example, the diagonal term corresponding to 
 \begin{align}
 \proj{01\cdots 0}
 \end{align}
where the projector contains $m$ 1's and $N-m$ 0's, will have a corresponding weight
\begin{align}
c^{2(N-m)}s^{2m}\left[\frac{1+\alpha}{2}\right]^{N-m}\left[\frac{1-\alpha}{2}\right]^{m}.
\end{align}
For the second term of \eqref{fourterms} we will have a similar situation, this time contributing 
\begin{align}
c^{2(N-m)}s^{2m}\left[\frac{1+\alpha}{2}\right]^{m}\left[\frac{1-\alpha}{2}\right]^{N-m}
\end{align}
to the same diagonal element. Adding these two contributions, each diagonal entry of $\rho_F$ containing $m$ 1's and $N-m$ 0's will have weight
\begin{align}
\gamma(m)=&c^{2(N-m)}s^{2m}\bigg(\left[\frac{1+\alpha}{2}\right]^{N-m}\left[\frac{1-\alpha}{2}\right]^{m}\nonumber \\
&+\left[\frac{1+\alpha}{2}\right]^{m}\left[\frac{1-\alpha}{2}\right]^{N-m}\bigg).
\end{align}
Turning to the third and fourth terms of \eqref{fourterms} we see that the only nonzero contributions from $\rho_0$ correspond to the $N$-fold tensor products of $\ketbra{00}{11}$ and $\ketbra{11}{00}$ respectively. These contribute to $\rho_F$ the two off-diagonal terms
\begin{align}
&(\alpha c s)^N \ketbra{00\cdots0}{11\cdots1}\nonumber\\& (\alpha c s)^N \ketbra{11\cdots1}{00\cdots0}.
\end{align}
Hence, we have a $\rho_F$ of the form
\begin{align}
\rho_F=\left(  \begin{array}{cccccccc}
    \gamma(0) & ¥ & ¥ & ¥ & (\alpha c s)^N \\ 
    ¥ &  \gamma(1) & ¥  & ¥ & ¥  \\ 
    ¥ & ¥ & \Ddots & ¥ & ¥  \\ 
    ¥ & ¥ & ¥ & \gamma(N-1) & ¥ \\ 
    (\alpha c s)^N &  ¥ & ¥ & ¥ & \gamma(N) \\ 
  \end{array}
\right)\nonumber.
\end{align}
We now define the quantity 
\begin{align}
w_0=\sum_{j=1}^{n} \sqrt{c_j d_j},
\end{align}
where $c_j$ $d_j$ correspond to entries in \eqref{gmecon}. Calculating this for $\rho_F$ we obtain
\begin{align}
w_0&=\sum_{j=1}^{n} \sqrt{c_j d_j} \nonumber \\
&=\frac{1}{2}\sum_{m=0}^N {N\choose m}\sqrt{\gamma(m)\gamma(N-m)\nonumber} \\
&=\frac{1}{2}c^Ns^N\bigg(\sum_{m=0}^N {N\choose m}\left[\frac{1+\alpha}{2}\right]^{N-m}\left[\frac{1-\alpha}{2}\right]^{m} \nonumber \\
&\quad\quad+\sum_{m=0}^N {N\choose m}\left[\frac{1+\alpha}{2}\right]^{m}\left[\frac{1-\alpha}{2}\right]^{N-m}\bigg)\nonumber \\
&=c^Ns^N\left(\frac{1+\alpha}{2}+\frac{1-\alpha}{2}\right)^N\nonumber\\
&=c^Ns^N.
\end{align}
Due to the form of $\rho_F$, we see that $|z_i|-w_i$ can only be positive for $i=1$. We have
\begin{align}
w_1&=w_0-c_1d_1 \nonumber \\
&=c^N s^N -\sqrt{\gamma(0)\gamma(N)} \nonumber \\
&=c^N s^N\left(1-\left[\frac{1+\alpha}{2}\right]^N+\left[\frac{1-\alpha}{2}\right]^N\right).
\end{align}
We may now calculate
\begin{align}\label{conc}
|z_1|-w_1=c^N s^N\left[\alpha^N+\left[\frac{1+\alpha}{2}\right]^N+\left[\frac{1-\alpha}{2}\right]^N-1\right].\nonumber
\end{align}
Finally, to calculate $C(\rho_{A_1\cdots a_N})$ we need to calculate the normalisation $\Tr[\rho_F]$. This is given by 
\small\begin{align}\
\Tr[\rho_F]&=\sum_{m=0}^N{N\choose m}\gamma(m) \\
&=\left[c^2\frac{1+\alpha}{2}+s^2\frac{1-\alpha}{2}\right]^N+\left[s^2\frac{1+\alpha}{2}+c^2\frac{1-\alpha}{2}\right]^N\nonumber \\
&=\left[\frac{1}{2}(1+\alpha\cos 2\theta)\right]^N +\left[\frac{1}{2}(1-\alpha\cos 2\theta)\right]^N.\nonumber
\end{align}\normalsize

Combining this with \eqref{conc} and using $\cos\theta\sin\theta=\frac{1}{2}\sin 2\theta$, we arrive at \eqref{concurrence}. For $\rho_{A_1\cdots A_N}$ we thus have
\small\begin{align}
C(\rho_{A_1\cdots A_N})=\frac{2\sin^N 2\theta \left(\alpha^N+\left[\frac{1+\alpha}{2}\right]^N+\left[\frac{1-\alpha}{2}\right]^N-1\right)}{\left[1+\alpha\cos 2\theta\right]^N +\left[1-\alpha\cos 2\theta\right]^N}.\nonumber
\end{align}\normalsize

\section{Appendix C: Genuine multipartite hidden nonlocality}
Here we calculate the fidelity between $\rho_{\tan\theta}$ and the $N$-party GHZ state. The state $\rho_{\epsilon}$ is given by 
\begin{align}\label{rhoep}
\rho_{\epsilon}=\frac{G_\epsilon^{\otimes N}\rho_{\text{\tiny{GME}}}\,G_\epsilon^{\otimes N}}{\Tr[ G_\epsilon^{\otimes N}\rho_{\text{\tiny{GME}}}\,G_\epsilon^{\otimes N}]}.
\end{align}
with
\begin{align}\label{gfilter}
G_\epsilon=\epsilon \proj{0} + \proj{1}.
\end{align}
Note that since $G_{\epsilon}$ has no support on the $\proj{2}$ subspace, only the first term of \eqref{povmlocal} will survive the filter. We may therefore replace $\rho_{\text{\tiny{GME}}}$ in \eqref{rhoep} by $\rho_{A_1\cdots A_N}$. To make calculations easier, we begin by working with the unormalised state
\begin{align}
\tilde{\rho}_{\epsilon}=G_\epsilon^{\otimes N}\rho_{A_{1}\cdots A_N}\,G_\epsilon^{\otimes N}.
\end{align}

Since the filter \eqref{gfilter} is diagonal, $\tilde{\rho}_{\epsilon}$ will have the same structure as $\rho_{A_{1}\cdots A_N}$. It is easy to see that after the filter, a diagonal element which contains $m$ 1's and $N-m$ 0's picks up a factor of $\epsilon^{2(N-m)}$ whereas the two off-diagonal elements pick up each a factor of $\epsilon^N$. We now use the ansatz $\epsilon=\tan\theta$. With this we have 
\begin{align}
\tilde{\rho}_{\tan\theta}=s^{2N}\left(  \begin{array}{cccccccc}
    \gamma'(0) & ¥ & ¥ & ¥ & \alpha^N  \\ 
    ¥ &  \gamma'(1) & ¥  & ¥ & ¥  \\ 
    ¥ & ¥ & \Ddots & ¥ & ¥  \\ 
    ¥ & ¥ & ¥ & \gamma'(N-1) & ¥ \\ 
    \alpha^N &  ¥ & ¥ & ¥ & \gamma'(N) \\ 
  \end{array}
\right)\nonumber,
\end{align}
where
\begin{align}
\gamma'(m)&=\left[\frac{1+\alpha}{2}\right]^{N-m}\left[\frac{1-\alpha}{2}\right]^m\nonumber \\
&\quad+\left[\frac{1+\alpha}{2}\right]^{m}\left[\frac{1-\alpha}{2}\right]^{N-m}.
\end{align}
For this state we have $\Tr[\tilde{\rho}_{\tan\theta}]=2 s^{2N}$ and so after renormalising we obtain
\small\begin{align}
&{\rho}_{\tan\theta}=\\
&\frac{1}{2}\left(  \begin{array}{cccccccc}
    \left[\frac{1+\alpha}{2}\right]^{N}+\left[\frac{1-\alpha}{2}\right]^{N} & ¥ & ¥ & ¥ & \alpha^N  \\ 
    ¥ &  \ddots& ¥  & ¥ & ¥  \\ 
    ¥ & ¥ & \gamma'(m) & ¥ & ¥  \\ 
    ¥ & ¥ & ¥ &\ddots & ¥ \\ 
    \alpha^N &  ¥ & ¥ & ¥ & \left[\frac{1+\alpha}{2}\right]^{N}+\left[\frac{1-\alpha}{2}\right]^{N} \\ 
  \end{array}
\right)\nonumber.
\end{align}\normalsize
One can now easily see how this state can be made arbitrarily close to the GHZ state. Taking $\alpha$ close to zero forces the extreme diagonal and off diagonal elements to $\frac{1}{2}$ while forcing all others to zero. Making this quantitative, we compute the fidelity between ${\rho}_{\tan\theta}$ and the pure GHZ state $\ket{\text{GHZ}}=(\ket{0}^{\otimes N}+\ket{1}^{\otimes N})/\sqrt{2}$:
\begin{align}
&{\cal{F}}({\rho}_{\tan\theta},\proj{\text{GHZ}})=\bra{\text{GHZ}}\rho_{\tan \theta}\ket{\text{GHZ}}\nonumber \\
&\quad\quad=\frac{1}{2}\left[\alpha^N +\left(\frac{1+\alpha}{2}\right)^N+\left(\frac{1-\alpha}{2}\right)^N\right]
\end{align}
which tends to $1$ when $\alpha$ tends to $1$. 

\end{appendix}

\end{document}